\documentclass{svproc}
\usepackage{url}
\usepackage{multirow}
\usepackage{amsmath}

\usepackage{amsfonts}
\usepackage{enumitem}
\usepackage{subcaption}
\usepackage[colorlinks=true,linkcolor=blue,anchorcolor=black,citecolor=blue,filecolor=blue,menucolor=blue,runcolor=green,urlcolor=blue]{hyperref}
\usepackage{array}
\usepackage{shortlst}
\usepackage{setspace}
\usepackage[pdftex]{graphicx}
\pdfoutput=1
\usepackage{ifpdf}
\begin{document}
\mainmatter              % start of a contribution
\title{Design of   Fuzzy Logic Controller  for  Washing Machine
}
\titlerunning{Fuzzy Controller}  % abbreviated title (for running head)
\author{Kriti Dheerawat \inst{1}}
\authorrunning{Kriti Dheerawat et al.} % abbreviated author list (for running head)
\tocauthor{Ivar Ekeland, Roger Temam, Jeffrey Dean, David Grove,
Craig Chambers, Kim B. Bruce, and Elisa Bertino}
\institute{The Maharaja Sayajirao University of Baroda}

\maketitle              % typeset the title of the contribution

\begin{abstract}
Things are becoming more advanced as technology advances, and machines now perform the majority of the manual work. The most often used home appliance is the washing machine for cloths. In this paper, we used the Mamdani approach and created an algorithm based on multi-input multi-output. The algorithm is implemented in Python. The results of this simulation show that the washing machine provides better execution at a low computation cost.% We would like to encourage you to list your keywords within
% the abstract section using the \keywords{...} command.
\end{abstract}

\section{Introduction}

Fuzzy logic is one of the agents of Artificial Intelligence that alter and portray uncertain information by analyzing the degree to which the statement is true. It was introduced by Professor Lofti Zadeh at the University of California, USA, in 1965. A related concept was conceived by American philosopher M. Black in (1937)\cite{Black}. He introduced the term “linguistic variables” in the year (1973) which defined a variable as a fuzzy set.\cite{1973}
Many uses of fuzzy set theory have been developed over the years, including fuzzy logic controller (FLC), fuzzy clustering algorithms, fuzzy mathematical programming, and many others. \\
%One of the most widespread of these is the fuzzy logic controller.\\ Fuzzy Logic enables to deliver a level of reasoning when deals with uncertainties. 
\subsection{Literature Review}

In 1975, Mamdani and Assilian\cite{EHMAMDANI} have developed the first fuzzy logic controller. In 2012, Zhen and Feng\cite{2012zeng} designed a gas heater based on the behavior Model of a FLC.  Also, for a liquid level control application, a FLC was designed using MATLAB which was compared to a PID controller. The comparison analysis showed that overshoots and steady-state errors were greatly decreased by the fuzzy logic controllers \cite{2012}.\\ A fuzzy inference approach is used in 2007 to calculate wash time where a triangualr Membership function  was used to determine the turbidity\cite{2021}.  Pritesh Lohani suggested an improved washer controller microchip in 2009, with inputs for type of dirt, mass of clothes, and degree of dirtiness, and an output of wash time with 25 rules \cite{2009}.\\

\vspace{-0.3cm}

 \section{Prerequisites}
We require concepts for fuzzy logic controllers described below: \\
\textbf{Fuzzy Relation:}\\
Let $X_1,X_2,\cdots,X_n$ be non empty sets then a  fuzzy relation $R$ is a fuzzy subset of $X_1\times X_2 \times\cdots \times X_n$.\\
\textbf{Fuzzy Intersection:t-norm}\\
A fuzzy intersection/t-norm $i$ is a binary operation in a unit interval that satisfies at least following axioms for all $a,b,c,d\in[0,1]$
\begin{enumerate}[label=Axiom-i\arabic*:,leftmargin=2.5cm]
\item $i(a,1)=a$
\item $b\leq d$ implies $i(a,b)\leq i(a,d)$
\item $i(a,b)=i(b,a)$ 
\item $i(a,i(b,d))=i(i(a,b),d)$
\end{enumerate}
\textbf{Fuzzy Logic Controller:}\\
A general fuzzy controller consists following steps:
\begin{enumerate}[label=Step-\arabic*:,leftmargin=1.50cm]
\item Initially we identify input and output variables of the controller and ranges of their values, after that we have to select meaningful linguistic states for each variable and express them by appropriate fuzzy sets.
\item Fuzzification :In this step a fuzzificztion function is defined for each input variable to express the associated measurement uncertainty . The following is the fuzzification function for the input variable 'a': 
\begin{center}    $f_{a}:[-b,b] \rightarrow F(\mathbb{R} )$,
   \end{center}
where $F:\mathbb{R} \rightarrow [0,1]$ and $f_a(x_0)$ is a fuzzy number for any $x_0\in[-b,b]$.
\item  Fuzzy logic control take decision using IF-THEN fuzzy inference rules. First, we convert given fuzzy inference rules of the form:
\begin{center}
If x is A and y is B and z is C Then w is W      
%If Type of dirt is Silt and Mass of cloth load is Medium and Type of cloth is Thin Then Wash Time is Short
 \end{center}
into fuzzy condition propostions  of the form
\begin{equation}\label{eq}
\text{If}\; (x,y,z)\; \text{is} \;A \times B \times C\; \text{then}\; w \;\text{is}\; W
\end{equation}
where A,B,C and W are fuzzy numbers choosen from the set of fuzzy numbers that represents the linguistic states and $X$, $Y$ and $Z$ are universal sets.
\item The inference engine evaluates the  rules contained in the fuzzy rule base utilising fuzzified measurements, and a fuzzified output is produced as a result.
For measurements input variables we first calculate degree of compability $R(x,y,z,w) $ by (\ref{eq1}) 
\begin{equation}\label{eq1}
R(x,y,z,w)=i_1[i_2(A(x),B(y),C(z)),W(w)]
\end{equation}
where $i_1,i_2$ are $t$-norms. The membership value for the combined control action W' is therefore given by 
\begin{equation}\label{eq2}
W'(w)=\bigcup_j sup_{\tiny{(x,y,z)\in X\times Y \times Z}}min\{ [f_{A}(x_0)\times f_{B}(y_0) \times f_{C}(z_0)], R_j(x,y,z,w)\}
\end{equation}
$\bigcup$ taken for union and $f_{A}(x_0), f_{B}(y_0)$, and $ f_{C}(z_0)$  are the fuzzy numbers for input variables $x_0,y_0,z_0$ respectively, defined as:
\begin{center}    $f_{A}:X \rightarrow F( \mathbb{R} )$,
      \;\; $f_{B}:Y \rightarrow F( \mathbb{R} )$,\;\;
 $f_{C}:Z \rightarrow F( \mathbb{R} )$,
\end{center}
\item In this last step of the design process, the designer of a fuzzy controller must select a sutaible defuzzification method for convert fuzzy output into  crisp value\cite{GeoK}.
\end{enumerate}

\section{ Proposed Design for Fuzzy Washing Machine}
Based on inputs and outputs, fuzzy logic is used to model a system.  The reduction of computational costs is the ideal factor of a smart washing machine. 
 
Following are  the steps for desiging FLC for washing Machine\\
Step-1: We identify ranges of input and outout variables.  
Step-2 : %The following are the fuzzification function for the measurement of input variables: 
%\begin{center}    $f_{TD}:X \rightarrow F( \mathbb{R} )$,
     % \;\; $f_{MC}:Y \rightarrow F( \mathbb{R} )$,\;\;
 %$f_{TC}:Z \rightarrow F( \mathbb{R} )$,
%\end{center}
%where $F:\mathbb{R} \rightarrow [0,1]$ is a fuzzy number.\\
The fuzzified inputs are measured by $f_{A}(x_0)$, $f_{B}(y_0)$, and $ f_{C}(z_0)$ for each $x_0\in X, y_0\in Y$ and  $z_0 \in Z$.  \\

Step-3: We select the linguistic inference rules. 
     This rules  are interpreted as disjuctive.

Step-4: For each rule, we first calculate degree of compability $R_j(x,y,z,w) $ by (\ref{eq1}), where minimum operator is used for t-norm.
\begin{equation}\label{eq4}
R_j(x,y,z,w)=min\{min\{(A_j(x),B_j(y),C_j(z)\},W_j(z))\}
\end{equation}
\vspace{-0.1cm}
 The membership value for the combined control action W' is therefore obtained using eq(\ref{eq2}) and given by (\ref{eq5})
\begin{equation}\label{eq5}
W'(w)=\bigcup_j sup_{\tiny{(x,y,z)\in X\times Y \times Z}}min\{[f_{A}(x_0)\times f_{B}(y_0) \times f_{C}(z_0)], R_j(x,y,z,w)\}
\end{equation}
Simmilarly  for each output we can evaluate fuzzy rule \\

Step-5:The fuzzified outputs wash time obtained from the fuzzy inference engine are converted to crisp values using the centroid method of defuzzification.
The overall area of the membership function distribution used to describe the combined control action W is subdivided into several predefined sub-areas, each of whose area and centre of area can be calculated.The following expression can be used to determine the controller's crisp output.
\begin{equation}
\dfrac{\sum_{i=1}^K A_if_i}{\sum_{i=1}^K A_i}
\end{equation}
where $K$ denotes the number of small areas or regions, $A_i$ and $f_i$ represent the area
and center of area, respectively, of $i$-th small region.
\vspace{-0.3cm}

\vspace{-0.3cm}
\section{Conclusion}
In this research work, the Mamdani inference approach have been used to design an algorithm for the fuzzy logic controller of a washing machine in Python programming language. We found that performance of the fuzzy controller is easier due to simplicity of the program structure.
The simulation results obtain after applying fuzzy controller shows that computation model makes the system more efficient and less expensive. It is clear that the type of dirt on the fabric, the thickness of the fabric, and the volume of the cloths all affect the amount of water, detergent, and wash time utilised.
The investigation shows that the suggested fuzzy controller improves the job efficacy of the washing machine with reducing requirement of water, detergent and wash time. Moreover by adding more aditional parameters in the system, one can optimise the performance of the machine as per requirement.
\section{Acknowledgement}
The authors express their sincere thanks to the reviewers  for their valuable suggestions for the improvement of the manuscript.

\end{document}